\documentstyle{elsart}
\input epsf

\begin{document}
\begin{frontmatter}
\title{Elemental Composition of Cosmic Rays near the Knee by Multiparameter Measurements
       of Air Showers}
\author[Chicago]{S. P. Swordy}
\author[Utah]{D. B. Kieda}
\address[Chicago]{Enrico Fermi Institute and Department of Physics, University of Chicago, Chicago IL 60637, U.S.A.}
\address[Utah] {Department of Physics, University of Utah, Salt Lake City, UT 84112, U.S.A.}
\
\begin{abstract}
The small change in the spectral slope of the overall intensity of cosmic rays near 1 PeV may be 
associated with the endpoint energy of supernova shock acceleration. A crucial test of this connection
and other ideas of the origin of the spectral `knee' is the reliable determination of the variation
of elemental composition in this region. Recent measurements at the DICE/CASA/MIA air shower
installation in Dugway, Utah, USA have provided several independent air shower parameters for each event.
The derivation of elemental composition from a combination of Cherenkov size, depth of shower
maximum in the atmosphere,
 muon size and electron size at ground level
and the reliability of these results are discussed. There
is no evidence from these data for a large change in the mean mass of cosmic rays across the
`knee'. These measurements show the cosmic rays are composed of $\sim 70$\% protons and $\alpha$ particles near total
energies of 10PeV.
\end{abstract}
\end{frontmatter}

\section{Introduction}
The slight steepening in the power law spectrum of the cosmic ray flux near energies of
$10^{15.5}$ eV has provoked a wide range of interpretations since its discovery in air shower
size spectra nearly 40 years ago\cite{kulikov}. This spectral `knee' remains an enigma, which together
with the more recently discovered flattening ( or `ankle') near $10^{18.5}$ eV provide the
only known features in the arriving cosmic ray spectrum at high energy. The coincidence of the
`knee' with the maximum energies expected from diffusive shock acceleration in supernova
remnants, has served to cement its value as a key region for further investigation in
cosmic-ray research. The features which might better connect the change in overall spectral slope
with expectations from shock acceleration lie in the details of the individual
elemental spectra hidden within the overall cosmic-ray flux. The accurate determination
of the elemental composition across the knee region, from $\sim 10^{14}$ to $10^{16}$eV,
is a crucial measurement yet to be made.
 
Recent observations of supernovae using TeV $\gamma$ rays have not managed to
detect \cite{buck2} the signal expected from high-energy cosmic-ray collisions with local
material if supernova remnants (SNR) are the sources of high-energy cosmic rays \cite{volk}.
Although EGRET has reported GeV $\gamma$ rays from some SNRs \cite{esposito},
it is still unclear if the signal is produced by locally intense cosmic rays \cite{brazier}. 
Non-thermal x-ray emissions have been detected from two SNRs \cite{koyama,keohane} 
which have been interpreted as synchrotron emission from local electrons at energies
up to $\sim$100 TeV \cite{reynolds}, but the source spectral slope inferred by these results seems
inconsistent with the electron source spectrum derived from direct measurements. 
It remains the case that the most convincing argument for supernovae explosions producing
the bulk of the cosmic rays arises from the staggering power required to maintain the
population given the measured lifetime of particles within our Galaxy. A typical value
for the power needed is $\sim 10^{40}$erg/s (see e.g. Schlickeiser\cite{schlickeiser94}),
based on a cosmic-ray residence time of $\sim 10^7$ years, 
a time-scale directly inferred from measurements of the surviving amount of the
radioactive isotopes of Be$^{10}$\cite{simpson} and Al$^{26}$\cite{simpsonal} which are created
by spallation reactions
of cosmic rays on interstellar material.

\section{Air Shower Detectors}
While direct detection above the atmosphere
is clearly the most desirable method for composition determination, the
fluxes of particles in this energy range are so small that a detector large
enough to collect a good statistical sample up to 10PeV (1PeV=$10^{15}$eV) has not yet been
built. Indirect detection of cosmic rays, through measurements of
air showers produced in the atmosphere, can easily supply sufficient
collecting power but these methods generally have poor
resolution for the properties of the originating cosmic ray nucleus
whose identity becomes clouded by the huge number of interactions in the
air shower. Further, the reconstruction of the mass of the incoming nucleus
from measurements of the shower distributions at ground level is 
subject to systematic errors introduced by reliance on shower
modeling by numerical simulations.

To address these issues we have constructed and operated a new type of
experiment designed specifically for cosmic-ray measurements near the knee.
This Dual Imaging Cherenkov Experiment (DICE)
is a ground based air shower detector which is designed to have
as little reliance as possible on the details of the air shower simulations
and to have the capability of comparison with existing direct measurements at
0.1PeV to provide an assessment of the overall systematic error.
DICE operates at the site of the 
Chicago Air Shower Array (CASA) and the MIchigan Array (MIA)
which are designed as an observatory for PeV $\gamma$ rays. They
can also be used to examine hadronic showers since they give measurements
of electron size and muon size at ground level.

\subsection{DICE description}
Since the method of imaging hadronic showers in Cherenkov light is a relatively recent development,
we provide some more detailed description of the DICE detectors and operation.
The two DICE telescopes are located at the CASA-MIA site in Dugway, Utah (described in Borione et al. (1994)\cite{borione94}).
They each consist of a 2m diameter f/1.16 spherical mirror with a focal plane detector of
256 close packed 40mm hexagonal photomultipliers (PMTs) which provide $\sim 1^0$ pixels in an overall
field of view $16^0 \times 13.5^0$ centered about the vertical.
The telescopes are on fixed mounts separated by 100m, (see Boothby et al. (1995, 1997)\cite{boothby95,boothby97}).

Cosmic-ray events within the field of view produce a focal plane image at the PMT
cluster which corresponds to the direction and intensity of Cherenkov light coming from
the air shower. When the direction of the air shower core and the distance of the shower
axis from the telescopes
are known, simple geometry can be used to reconstruct the amount of light received from each
altitude of the shower. The amount of Cherenkov light produced is strongly correlated with
the number of electrons in the shower and is used to estimate the electron size as a function
of depth in the atmosphere from which the location of shower maximum can be determined. This procedure
is essentially geometrical and is independent of numerical simulations except for calculations
which determine the angular distribution of Cherenkov light around the shower axis.

\subsection{CASA/MIA description}
The CASA/MIA installation and operation is described in
detail in Borione et al. 1994\cite{borione94}. In this present work,
 information from CASA-MIA is used to establish the
event geometry and to directly measure the electron size and muon size at ground level. Since the event
acceptance is constrained by the DICE aperture, only events within $\sim$10$^0$ of the zenith
are analyzed, making the atmospheric depth of the size measurements to be essentially the
same for all events at $\sim$860g/cm$^2$.

\section{Measured Parameters}
For each air shower collected a simple time coincidence is used to identify the same event
in both DICE clusters and CASA-MIA. Further requirements on the correlation of the DICE images
with each other and with the CASA-MIA event geometry are used to reduce the overall probability
of event mismatches between the detectors to $\sim$10$^{-5}$.
The basic independent parameters measured for each shower
are shown in Table \ref{parms}

\begin{table}
\caption{\bf Parameters measured for each event.\label{parms}}
\vspace{0.5cm}
\begin{center}
\begin{tabular}{|l|l|l|}
\hline
Description &   Symbol used here & Source Instrument \\
\hline
  Shower direction and core location &  & CASA \\
  Cherenkov size & $Ch$ & DICE\\
  Shower maximum & $X_{max}$ & DICE\\
  Electron size at ground & $N_e$ & CASA\\
  Muon size at ground & $N_{\mu}$ & MIA \\
\hline
\end{tabular}
\end{center}
\end{table}

The parameters for each shower are derived from these measured values. The accuracy of
the shower core location derived by CASA is 1-3m depending on the overall shower size.
The measurement of the shower arrival direction is accurate to $\sim 0.4^0$ for larger showers
with some degradation for lower energy events. These measurements and their errors are
discussed in detail in Borione et al. 1994\cite{borione94}. The electron size from CASA is derived
from a fit to the measured particle intensities to signals produced in an array of plastic
scintillation counters using the NKG formulism\cite{greisen60,kamata58}. The muon
size is a fit to signals produced in buried scintillators using 
the function discussed by Greisen\cite{greisen60}. Further details of these
fits are discussed in McKay\cite{mckay92}. 

\subsection{Derived Shower Parameters}
Some of the shower parameters used in this analysis correspond directly to measured values.
The electron and muon sizes at ground level ($N_e$ and $N_\mu$) are those derived by CASA/MIA.
The Cherenkov size at the two DICE detectors is obtained
by summing the total amount of light detected at each photomultiplier tube (PMT) cluster.
The location of shower maximum in the atmosphere ($X_{max}$) is determined by fitting the shape
of the shower image in each of the DICE PMT clusters.
The Cherenkov light from each height in the shower
falls into a specific
angular bin in the focal plane when the event geometry is known.
By applying corrections for solid angle, the variation of light emission intensity with respect to the 
shower axis, 
and
corrections for atmospheric absorption, a profile of electron size versus atmospheric depth is developed
from the Cherenkov image. This is fit with a simple two parameter shower shape to find the location of shower
maximum in each DICE telescope.
 An estimate for ($X_{max}$) is found for both DICE clusters,
a crucial difference between DICE and detectors which fit the lateral distribution of Cherenkov
light is that this estimate of ($X_{max}$) is based on simple geometry. It does not rely on
the details of numerical simulations or have stringent linearity requirements of the
detectors.
 
Previous work with DICE estimated the shower energy 
by a simple translation from the total amount of Cherenkov light
in the image and the geometry of the shower\cite{bapjl97}. In this present work a more accurate estimate
of energy is derived from a combination of the amount of Cherenkov light and the $X_{max}$ determination 
produced by each DICE telescope. 
This is desirable since the lateral distribution and intensity of Cherenkov light at a given
total energy depends both on the primary particle mass, hence mean $X_{max}$, and the distance of the measurement from the
shower core.
A fit for the total shower energy and primary particle mass is made to the geometry, Cherenkov size ($Ch$)
and $X_{max}$ location in the two DICE telescopes.
The form of the Cherenkov size function used in these fits is derived from the results of simulations using the program
CORSIKA 4.50 with the VENUS interaction model(\cite{corsika98}).
The derived $X_{max}$ fitting function has a constant shower elongation
rate parameter and assumes simple superposition for providing primary particle mass dependence.
The exact form of both of these fitting functions are given in Appendix A as equations \ref{eq:csize} and
\ref{eq:xmax}.

\begin{figure}
\vspace{-0.5cm}
\hbox to \textwidth{\hss
\epsfxsize=15cm
\epsfbox{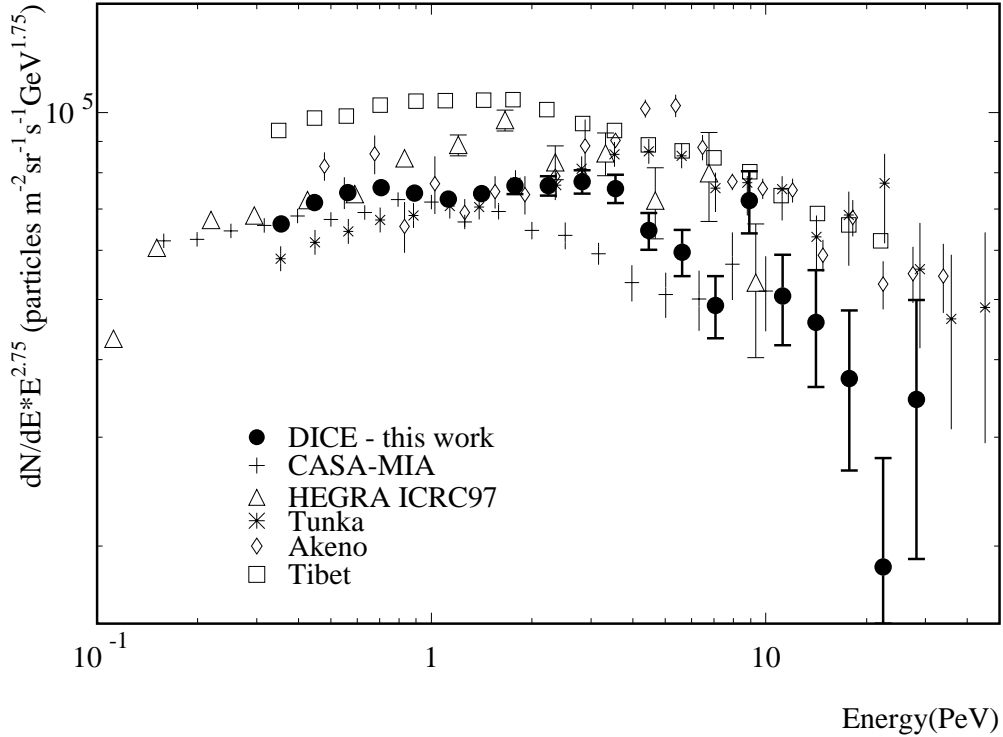}
\hss}
\caption{Measurements of energy spectra near the `knee' from this work (filled circles) compared with recent
measurements from Tibet\cite{tibet96}, CASA-MIA\cite{casa99}, HEGRA\cite{hegra97}, 
Tunka\cite{tunka97}, and Akeno\cite{akeno84}.
\label{espec} }
\end{figure}

Events collected by DICE in coincidence with CASA/MIA over a period from mid 1994 to early 1996 are
subjected to the following selection cuts before fitting with equations \ref{eq:csize} and \ref{eq:xmax}.
(1) The core of the shower lies at a distance $100m < r < 225m$ from both DICE telescopes. (2) The fits of the
longitudinal development in both DICE telescopes have reduced $\chi^2 < 3$. (3) The $X_{max}$ from each
telescope agrees within 150$g/cm^2$. (4) The arrival direction of the shower is within 6$^0$ of the vertical.
The fits to the measured $Ch$ and $X_{max}$ from the two sites are made using the minimum $\chi^2$
method.   
With these cuts the effective collection
geometry is determined by the instrument Monte Carlo to be $\sim$3300 m$^2$sr,
making the overall collecting power $\sim$125,000 m$^2$sr days.
Using the detailed 
calculated aperture and efficiency corrections derived from an instrument Monte Carlo discussed below
the cosmic ray energy spectrum can be constructed from the fit energies.
 This is shown as the filled circles in
Figure \ref{espec} together with previous determinations of the energy spectra in this region.
The fluxes have been multiplied by $E^{2.75}$ to emphasize the change in slope across the knee
region. 
These agree with the previously reported energy spectrum for DICE \cite{bapjl97}.
We believe this new analysis provides a better determination of fluxes because of the
primary mass dependent
fitting function used for each event. 
These new DICE data are given in tabular form in Appendix B. The `knee' here is at an energy
around 3PeV.

\begin{figure}
\vspace{-0.5cm}
\hbox to \textwidth{\hss
\epsfxsize=15cm
\epsfbox{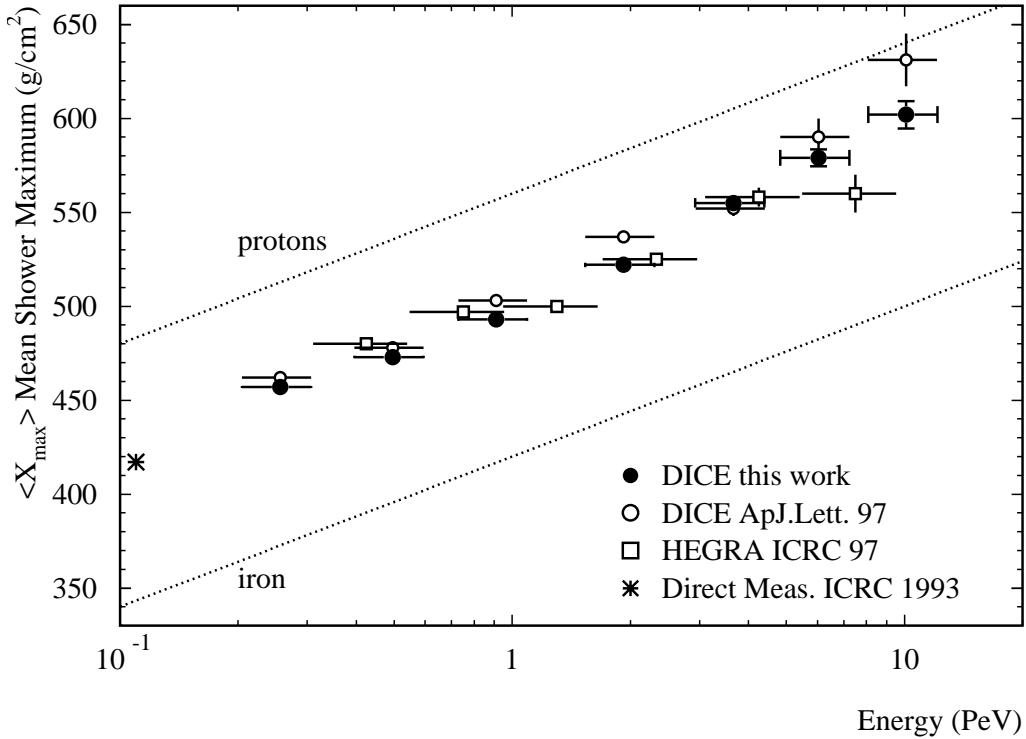}
\hss}
\caption{Measurements of mean  $X_{max}$ versus Energy 
\label{xmax} }
\end{figure}

The mean measured values of $X_{max}$ as a function of energy is shown in Figure \ref{xmax}. Here these values
are compared with the previous DICE results\cite{bapjl97} and those of the HEGRA experiment determination\cite{hegra97} which
uses the distribution of Cherenkov light at ground level to estimate $X_{max}$.
The star shows the expectation for
the mean $X_{max}$ based on a the direct composition measured near 100TeV\cite{sword93} using
equation \ref{eq:xmax}.
The various measurements show general agreement, although there is a tendency for
the DICE results to show a lighter composition at the highest energy. The present data are in statistical agreement with
the previously published results of DICE, but they show somewhat less departure
from constant composition at high energies. The dashed lines are for pure proton or pure iron
composition given by equation \ref{eq:xmax}. These results are tabulated in Appendix B.

\subsection{Robustness of Shower Parameters}
The determination of incident particle mass from air shower measurements have generally
poor resolution. Consequently features such as elemental peaks, which can commonly be used in direct
measurements of cosmic rays to test detector calibration, are absent. This makes systematic 
sources of error the largest obstacle which must be overcome in order to produce a reliable
measurement. Any internal consistency checks which can be used on the data are valuable as a way
to keep these problems to a minimum. The combination of DICE and CASA/MIA provides a unique combination
of shower parameters which over-determine the characteristics of each shower. The comparison of 
parameters from different detectors can be used to test the reliability of results. This process
is already possible within DICE since the separate measurements of $X_{max}$ and $Ch$ made with
each cluster can be compared to test consistency. 
A more rigorous test is a comparison of the information derived
from DICE with that from CASA/MIA.

The general profile of air showers produced by hadrons has been studied by several authors.
These showers can be thought of as a succession of electromagnetic showers fed by a core of
interacting hadrons. The fluctuations in shower maximum are produced both by the statistical
nature of the shower development and the fluctuations in the first interaction point in the
atmosphere. At atmospheric depths deeper than $X_{max}$, the shower size declines in a manner
which is dominated by the atmospheric hadronic interaction length. If the size is measured
at a depth relative to the location of $X_{max}$ the fluctuations produced by variations
in the initial interaction point are removed. We can perform a simple test on the data
discussed here: Does the electron size at ground level measured by CASA fit with the shower
energy and location of $X_{max}$ produced by DICE? 
By using the average shower development function quoted by Gaisser\cite{gaisser} (given in Appendix A),
we can compare the expected ground electron size with the CASA measurement. 

\begin{figure}
\vspace{-0.5cm}
\hbox to \textwidth{\hss
\epsfxsize=15cm
\epsfbox{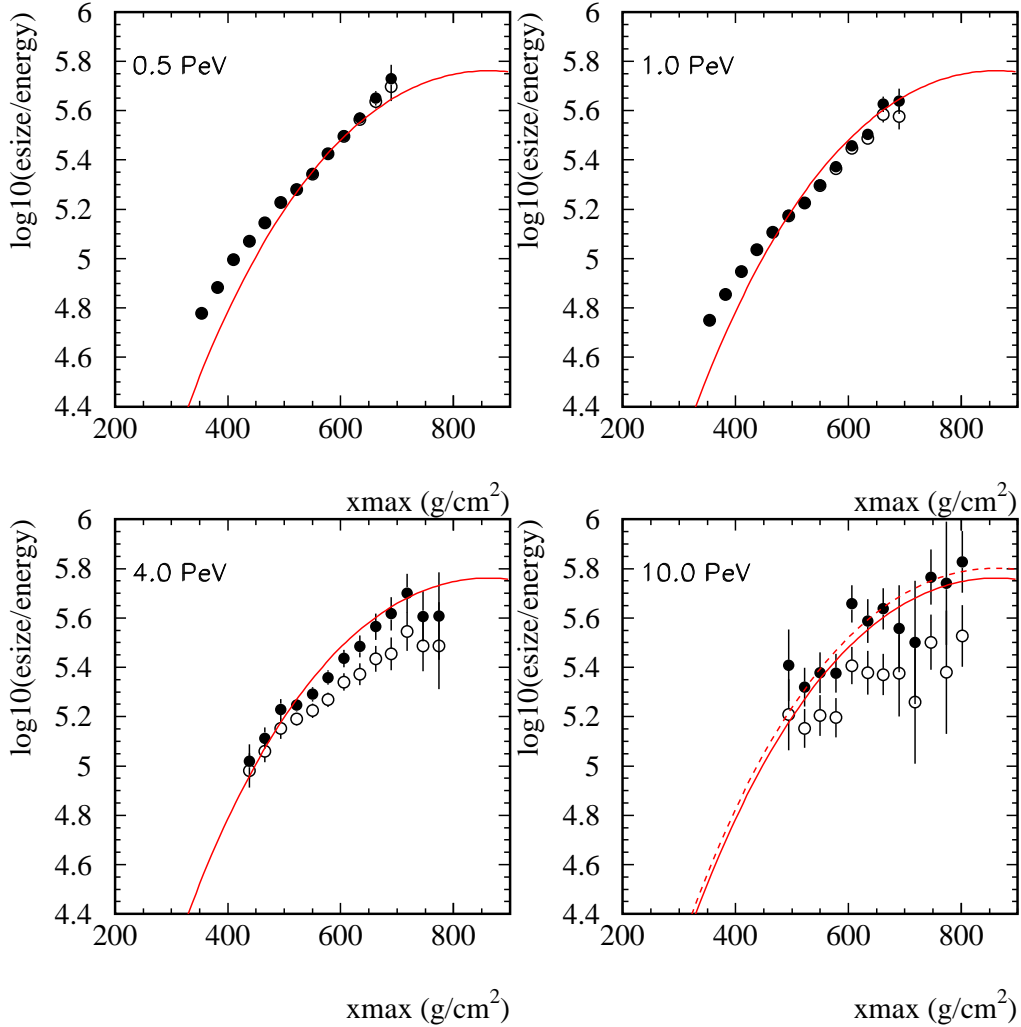}
\hss}
\caption{Comparisons of expected electron size at ground with a shower model from Gaisser \cite{gaisser}.
The open symbols represent raw electron sizes, the filled symbols are electron sizes corrected for saturation
in a manner discussed in the text. 
\label{gfig} }
\end{figure}

Since showers at a given energy have values of $X_{max}$ which fluctuate for the reasons outlined
above, we can sample a range of potential electron sizes versus $X_{max}$ for a specific primary
energy range. This procedure can be used to verify the consistency of measurements and to test
for systematic errors in the determination of $X_{max}$, $E$, and $N_e$ as a function of energy.
If at a ground level depth of $X_0$ we compare $N_e(X_0)/E$ with $X_{max}$ for each event 
we can test on average if these agree
with equation \ref{eq:gaiss} over a range of parameters. Figure \ref{gfig} shows how this test
works for $X_{max}$ and $E$ determined from DICE and $N_e$ from CASA. The four panels correspond
to some of the energy bins used for Figure \ref{xmax}. Each panel shows the average log10($N_e/E$)
plotted versus shower $X_{max}$. The function derived from the average air shower development 
(from equation \ref{eq:gaiss}) is shown as a solid line. The good agreement between the measurements
and the data for the lower energy data provide confidence that the location of $X_{max}$ determined
by DICE and the electron size at ground level are measured correctly. The version of the Gaisser function
shown as the dashed curve in the lower right hand panel of Figure \ref{gfig} shows the effect of scaling violation
as discussed in Appendix A.3.

It is important to realize
that not just the shower shape but the normalization of the shower size versus $X_{max}$ is
given by equation \ref{eq:gaiss}, there is no arbitrary vertical normalization of the curves
shown in Figure \ref{gfig}. These results therefore also provide confidence that the measurement
of energy from DICE is reasonable, an error in energy would produce a systematic shift of the
data in the vertical direction. For the higher energies there appears to be a small systematic
shift, of the order of 10\%, of the data to a region below the curve. This could be
produced by a various effects; (i) the energy scale is overestimated at the higher
energies, (ii) the electron size is underestimated for large electron sizes,
(iii) the location of $X_{max}$ is overestimated for high energies.
This last possibility seems unlikely since the method used for the measurement of $X_{max}$ is
the same at both low and high energies and no similar systematic shift appears across all energies.
For example the region for $X_{max}$ near 700g/cm$^2$ is sampled both by the 1PeV and 10PeV energy,
at the lower energy the measurements agree with the expected curve, at higher energies they do not.
A systematic error in the determination of $X_{max}$ for deep showers
should produce a similar shift at both energies.
The lack of a systematic shift across all energies also shows that errors
in $X_{max}$ produced by atmospheric scattering of Cherenkov light is minimal, since this would also
only depend on the depth of maximum shower size.
The first possibility, (i) mentioned above, also seems unlikely since there would be a large effect on the energy
spectrum shown in Figure \ref{espec}. A shift in the energy scale measure of the required magnitude ($\sim$ 20\%)
would  produce a significant increase in the apparent fluxes at the higher energies ($\sim$ 40\%),
which does not seem to be present in the data. We are left with the possibility that the electron
size somehow `saturates' slightly at the higher energies if we believe equation \ref{eq:gaiss} is
independent of energy. This is a possibility since CASA, primarily built for observations near 100TeV,
was not designed to operate for air shower core density measurements at 10PeV. The electron sizes
at high energy are predominantly derived from counters on the perimeter of the core which have
not saturated. We have developed a simple correction to the electron size for large values which
apparently corrects the discrepancy shown in Figure \ref{gfig}. This is given by:

\begin{eqnarray*}
log_{10}N_{ec}& = & log_{10}N_{e} \ \ \ \ log_{10}N_{e}<5.5 \\
log_{10}N_{ec}& = & log_{10}N_{e}\times (1+(log_{10}N_e-5.5)\times {0.04}) \ \ \ \  log_{10}N_{e}\ge 5.5 \\
\end{eqnarray*}

Where $N_{ec}$ is the corrected electron size, $N_{e}$ is the original size and $E$ is the shower energy
in PeV as described above.
This increases the electron size by around $\sim$ 7\% near energies of 10PeV. This simple parameterization
provides a much better fit to the shower curves in Figure \ref{gfig} as shown by the filled symbols
identified as `saturation corrected data'. We use this corrected electron size through the rest of this paper.
Importantly this change is not large, but illustrates the power of correlating apparently redundant
measurements. These can be used to explore systematic problems with various measured quantities.

\section{Mass Determination}
In this analysis we adopt a slightly different philosophy from previous work by using combinations
of measured parameters to derive an incident particle mass, $A$, on a shower by shower basis. With
the information presented above we form two estimates for the mass, one from the location of $X_{max}$
and the fitted shower energy and another from the muon and electron sizes in combination with the
fitted energy.

\subsection{Mass Functions}
Although mass functions of arbitrary complexity can in principle be derived from analysis of numerical
simulations, these are not always transparent as to the underlying physical processes which
produce the sensitivity to mass from a particular combination of parameters. Here we adopt the
approach of using a simple physical model to produce a mass function which is tested against
simulations to determine the accuracy of this function over some range of energies. We first make
some simple assumptions about the variation of shower parameters for singly charged
particles with energy:

\begin{eqnarray*}
X_{max} & = & X_0 + X_m \times log_{10}(E) \\   
N_{\mu} & = & N_{\mu0} \times E^\alpha \\
\end{eqnarray*}

Where $X_0$, $X_m$, $N_{\mu 0}$, $\alpha$, are constants and $E$ is the
energy in PeV. Typical values for the scaling constants are $X_m=80g/cm^2$ and
$\alpha=0.87$.
We assume simple superposition to define how these will vary for heavier
particles of mass A:

\begin{eqnarray*}
X_{max} & = & X_0 + X_m \times log_{10}(E/A) \\
\mu_{g} & = & log_{10}(A)+\alpha \times log_{10}(E/A) \\
\end{eqnarray*}

Where $\mu_g=log_{10}(N_{\mu}/N_{\mu 0})$ and the first of these equations is the same as \ref{eq:xmax}.
We can rearrange these to give an estimate for the mass, A:

\begin{eqnarray}
log_{10}(A)_x & = & (X_0 - X_{max})/X_m + log_{10}(E)  \label{eq:xmass} \\
log_{10}(A)_g & = & (\mu_g - \alpha\times log_{10}(E))/(1-\alpha)  \label{eq:mumass} 
\end{eqnarray}

Where $log_{10}(A)_x$ is the mass from the measurement of $X_{max}$ and $log_{10}(A)_g$ is the mass from the measurement
of muon size at ground level. At fixed energy, deeper showers (larger $X_{max}$) give a smaller
mass and muon rich showers (larger $\mu_g$) give a larger mass. Another feature of these mass
estimate equations is that they have an opposite sign dependence on the shower energy. Hence
errors in energy assignment produce opposing errors in the mass estimates. In principle these
estimates can be used to make independent mass estimates from the same showers. Equation \ref{eq:xmass}
results from the location of the hadronic shower maximum and equation \ref{eq:mumass} arises
from the energy of the pions decaying into muons in the shower. At the same total energy, proton
showers have typically higher energy pions than iron showers so heavier nuclei produce more muons
from pion decays in the shower development. 
  
There is however another factor which effects the muon size at ground level. This is slightly
dependent on $X_{max}$ in a similar but much weaker fashion to the electron size dependency
discussed above and shown in Figure \ref{gfig}. Fluctuations in shower development produce
larger muon sizes at ground level when $X_{max}$ is larger for the same initial energy.
Although this is not a large effect, the mass estimate described in equation \ref{eq:mumass} is sensitive
to small fluctuations in $\mu_g$ because of the small value of $(1-\alpha)$ in the denominator.
These fluctuations are strongly correlated with the electron size at ground level and their effect
can be significantly reduced by introducing a correction term based on $e_g=ln(N_{ec}/N_{e0})$, the
normalized electron size at ground level.
The overall mass function derived from muon measurements using this correction is given by:

\begin{eqnarray}
log_{10}(A)_g & = & (\mu_g - \alpha\times log_{10}(E) - \delta(e_g))/(1-\alpha)  \label{eq:muemass}
\end{eqnarray}

Where the correction $\delta(e_g)$ is applied to reduce fluctuations in muon size arising from
variations in the shower development.
This correction is discussed in detail in Appendix C.
 
\subsection{Correlations of Mass Estimates}
\begin{figure}
\vspace{-0.5cm}
\hbox to \textwidth{\hss
\epsfxsize=15cm
\epsfbox{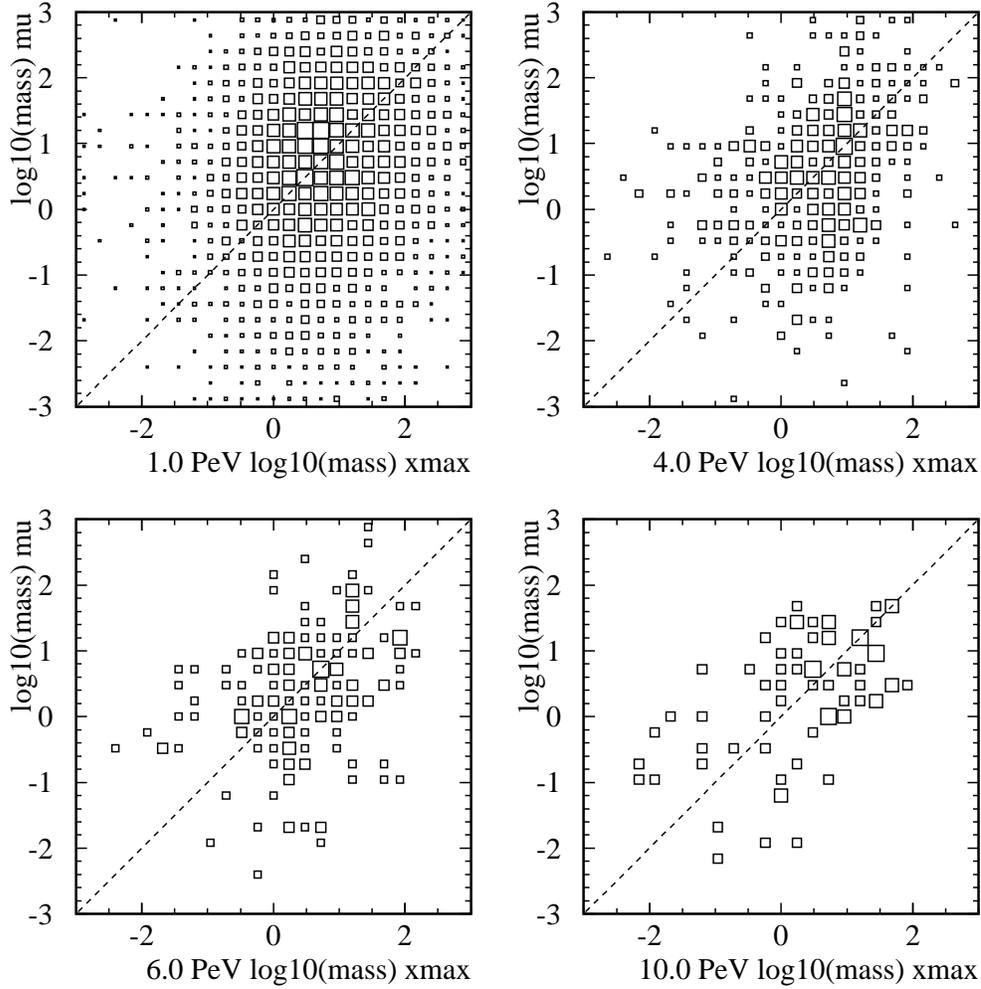}
\hss}
\caption{Cross correlation plots of the mass determinations from $X_{max}$ and the muon size for
events in energy bins similar to Figure \ref{xmax}. The size of each box is linearly proportional to
the number of events, concentration of the events are expected along the dashed lines if the correlation
is good. The mass functions are discussed in the text. 
\label{crossmass} }
\end{figure}

To investigate the consistency of these mass estimates we can look at the variations of one type
of mass estimate versus the other. This is shown in Figure \ref{crossmass} for some of the energy bins studied.
The correlation between these estimates over a range of energies provides confidence that the
estimates are roughly correct. This correlation has some small contribution from the common energy term in
equations \ref{eq:xmass} and \ref{eq:mumass}, this is not large since a fluctuation in energy of 30\%
only produces a change in either mass function of $\sim$ 0.1. The dominant contribution to this correlation
is from the rapid variation of electron size measured at the ground with $X_{max}$ depth as shown in
Figure \ref{gfig}. Although clearly $\mu_{g}$ plays a role, since there are more muons from 
showers produced by heavier primaries, the fluctuations of showers at some fixed primary energy
and mass provide wide enough variations in $X_{max}$ to make the electron size effect predominate. In some sense
the correlations in Figure \ref{crossmass} are a straightforward result of the relationships shown in
Figure \ref{gfig}. 

\subsection{Comparison with Simulations}
To further test these distributions and methods for estimating mass we can compare these results
with the expectations from simulations. These calculations must take account of shower properties,
detector properties, and the analysis cuts used. A simulation program 
which uses the results of air showers produced by the CORSIKA 4.5 code as an input for the shower
characteristics. 
The simulation effort was concentrated on the DICE measurements of $X_{max}$, all the details of
the fluctuations in the quantities directly affecting these were included. The electron sizes and
muon sizes at ground level are included but the fluctuations in these are parameterized in a simplified
fashion to speed up the calculations.  
These simulated showers impact on the detector site at random locations and angles near
the vertical in the simulation. The output of the simulation is a data file which can be subjected
to exactly the same cuts and analysis procedures applied to the actual data. 
For the analysis discussed here the input primary mass composition used in the simulation is
a constant 30\% H, 30\% He, 10\%CNO, 10\% NeMgSi, and 20\% Fe. As discussed later, there is no
evidence for significant variation in this composition across the range of energies in this work.
The simulation is also generated with a spectral shape which approximates the variation in
slope through the knee region.
and includes details of the contributions from shower geometry, focal plane pixel size, etc.
 The simulation can also be used to determine the overall effective
geometrical factor of the instrument. This was used in combination with the instrument livetime
to calculate the normalization of the data in Figure \ref{espec}.

\begin{figure}
\vspace{-0.5cm}
\hbox to \textwidth{\hss
\epsfxsize=15cm
\epsfbox{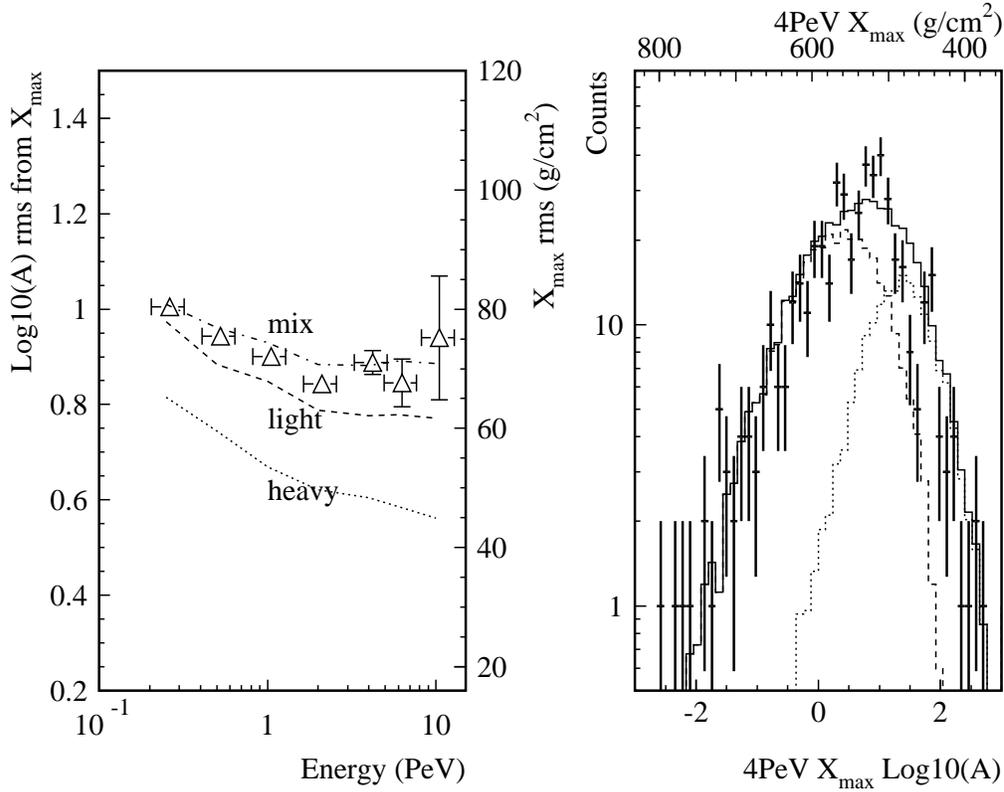}
\hss}
\caption{(Left) $X_{max}$ mass distribution RMS widths compared with simulations across all measured energy bins. The right hand axis is labelled with the equivalent RMS width in $X_{max}$.
(Right) Detailed $X_{max}$ mass distribution at 4PeV compared with simulations discussed in the text.
The top axis is labelled with the corresponding values of $X_{max}$ using equation \ref{eq:xmax}
\label{masswidth} }
\end{figure}

We can compare the shape of the $X_{max}$ mass estimates from DICE 
with the predictions from the simulation.
This provides a test of the overall consistency of the method used and can exclude certain
extreme composition possibilities.
The left hand panel of Figure \ref{masswidth} shows the RMS value of $log_{10}(A)$ compared with expectations
from simulations for mixed composition discussed above, light (p+$\alpha$), and heavy (A$>$4)
composition.
The comparison of the complete distribution for the $X_{max}$ mass at 4PeV
is shown in the right panel of Figure \ref{masswidth} as data points with errors.
This also shows the expected distributions for
a mixed composition (line), light (dashes), and heavy (dots). Although this is not the method of choice
for the most accurate composition determination, these data alone seem to exclude a pure Fe
composition over all of this energy range.
Unlike other methods which use the distribution
of Cherenkov light at ground level, the mass determination of DICE does not produce a long tail for
$log_{10}(A)_x$ in the negative region. This is because DICE intrinsically rejects all showers with
an apparent $X_{max}$ deeper than $\sim 750g/cm^2$ by geometry. This provides some truncation of the
distribution produced of the mass function which is incorporated in the simulation.

\begin{figure}
\vspace{-0.5cm}
\hbox to \textwidth{\hss
\epsfxsize=15cm
\epsfbox{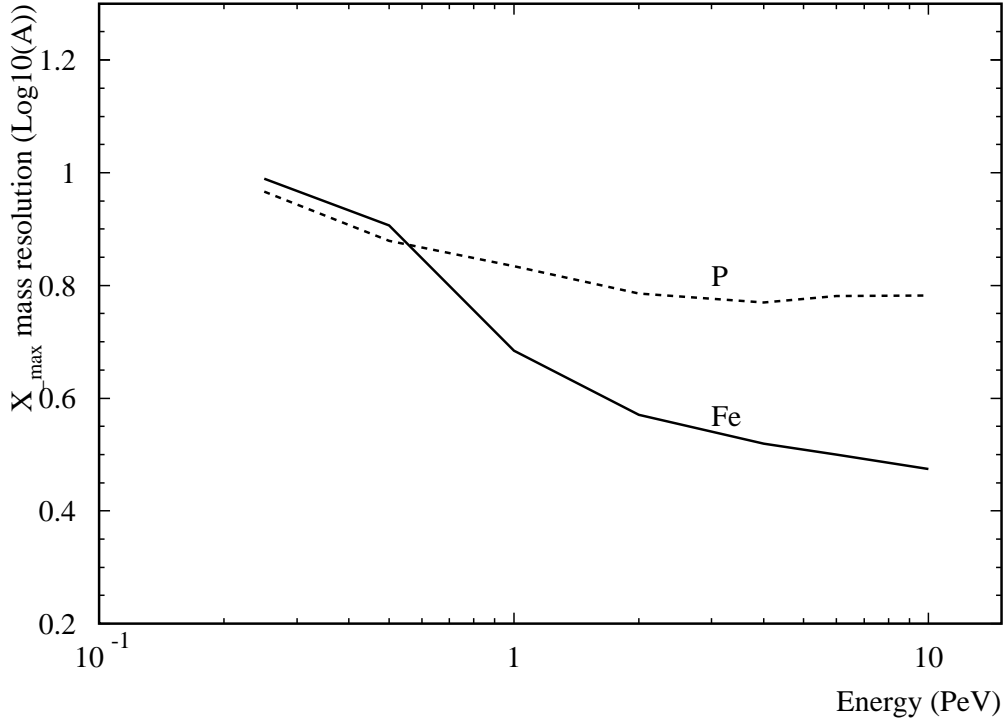}
\hss}
\caption{The RMS resolution derived from simulations
for single events for $log_{10}(A)$ derived from the measurement of $X_{max}$ by DICE.
The dashed curve is for proton events, the solid line for iron nuclei.  
\label{resolution} }
\end{figure}

With distributions such as shown in the right hand panel of Figure \ref{masswidth} the mass resolution of the
DICE $log_{10}(A)_x$ measurements can be estimated. These are shown in Figure \ref{resolution} for incident protons and
iron nuclei as a function of particle energy. The ordinate shows the RMS resolution of the DICE experiment for
the $log_{10}(A)_x$ of a single event. The improvement in the resolution with increasing energy arises from a
combination of larger light yield and an improving $X_{max}$ resolution with the better event geometry
in DICE of deeper showers. The mass resolution of proton showers at energies above a few PeV is close to the
intrinsic limit for a mass based on $X_{max}$ determination. Even if the depth of shower maximum and the shower
energy is known perfectly, the mass resolution in $log_{10}(A)$ for protons is limited to $\sigma\sim$0.8 because of intrinsic
fluctuations in the shower development. 
The mass resolution for iron showers is significantly better than for protons because the intrinsic fluctuations
of showers from these nuclei are significantly smaller.
It is sobering to realize that iron is separated by only $\sim 2\sigma$ from an incoming proton signal for single events.

\section{Composition Determination}
\begin{figure}
\vspace{-0.5cm}
\hbox to \textwidth{\hss
\epsfxsize=15cm
\epsfbox{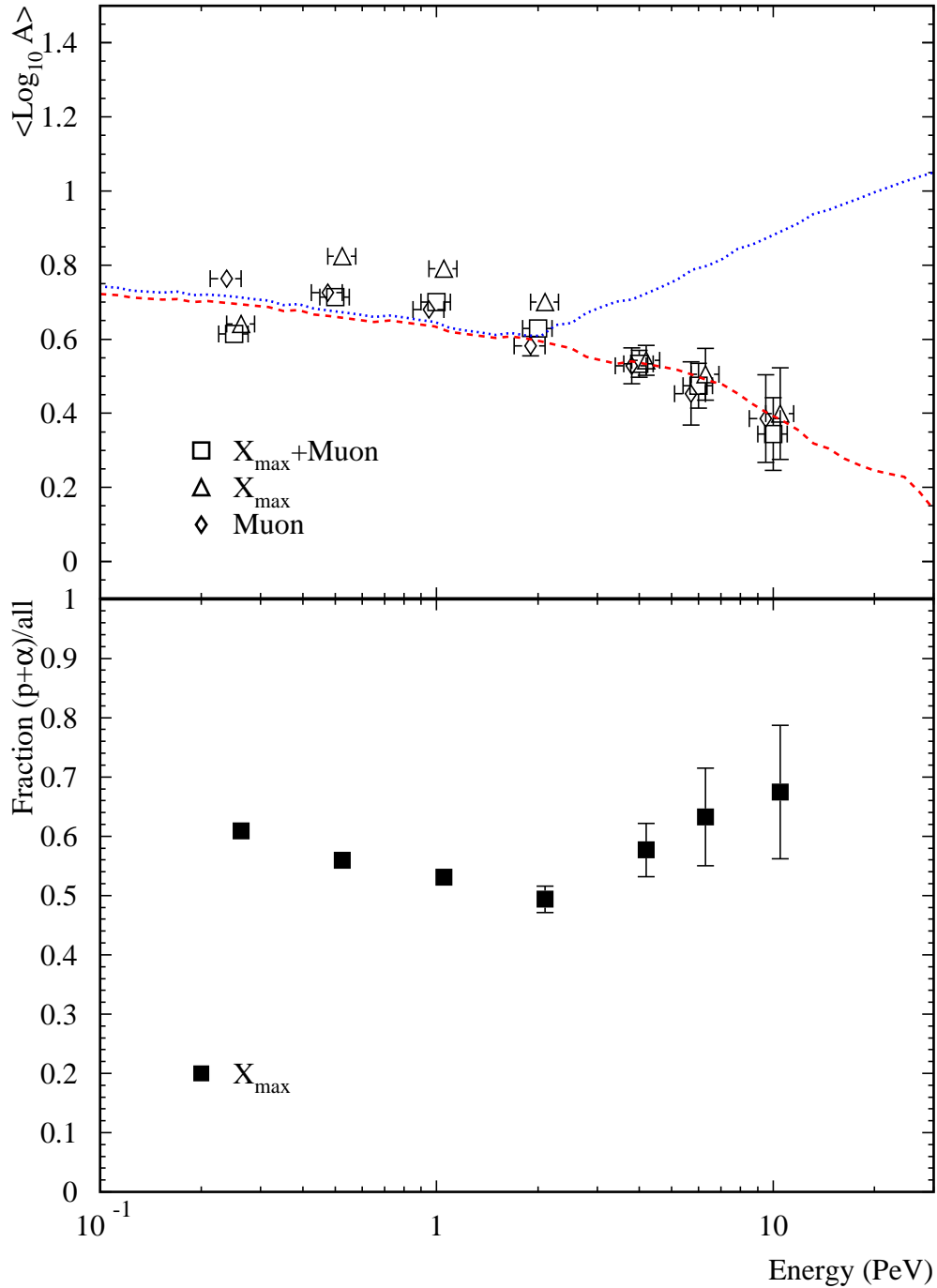}
\hss}
\caption{(Top) Results for mean $log_{10}(A)$ measurements from this work versus energy, the model curves are
discussed in the text. (Bottom) Derived fraction of (p+$\alpha$)/all for the $X_{max}$ mass determination.  
\label{massplot} }
\end{figure}

Using these methods a mass estimate 
for each event can be derived from (i) $X_{max}$ , (ii) muon and electron
size, and (iii) a simple linear combination of the two, called $log_{10}(A)_c$.
This last method has the advantage of somewhat increased resolution because the methods are to some extent independent.
Each event is assigned values for $log_{10}(A)_x$, $log_{10}(A)_g$, and $log_{10}(A)_c$ 
and collected in bins of energy. 
The results for the mean $log_{10}(A)$ values versus energy are shown as the upper panel
in Figure \ref{massplot}. 
The mass error is based on statistics and the energy error is the estimate of the systematic error in the energy scale
of the fits discussed earlier. All of the mass estimates show variations with energy across
the knee region. These data are tabulated in Appendix B.

Another method for determining composition has been used (see e.g. Watson\cite{watson}) which attempts to separate a
sample of almost pure proton+$\alpha$ events by applying a simple cut to the mass distribution,
such as shown in Figure \ref{masswidth}, which excludes heavy particles. The simulation can then
be used to make corrections to the samples produced by estimating the amount of `leakage'
of heavier events into the (proton+$\alpha$) sample. The results of an analysis on these data
which uses a cut to extract the events where $log_{10}(A)\le 0$ are shown in the lower panel of Figure \ref{massplot}
for the DICE $X_{max}$ results. Similar analyses have not been made for the ground size distributions because
simplifications in the fluctuations for these data make this method unreliable.
The errors on the (p+$\alpha$) fraction are based on simple counting
statistics, the uncertainty is larger for this analysis method because fewer events pass the data cuts.
There is a tendency of these results
to a higher fraction of (p+$\alpha$) at the higher energies although they are also consistent with constant
composition across this region. The values of the (p+$\alpha$) fraction are tabulated in Appendix B.

\section{Discussion}
By using several parameters measured for each shower a mass estimate for each incoming primary
particle can be extracted which shows remarkable consistency between values derived from $X_{max}$
and from the $\mu/e$ method. The direct observation of the expected shape and absolute normalization of
showers in this energy range (Figure \ref{gfig}) is a crucial test for the absence of systematic errors
in the methods of mass estimation. The widths of the distributions shown in Figure \ref{masswidth} are
consistent with the results shown Figure \ref{massplot}.
These results essentially exclude an all-iron composition at 10 PeV, this
conclusion was reached by an earlier analysis of DICE.
The results suggest a tendency to lighter composition at high energies,
although those from the ratio method are consistent with constant composition.
A composite mean $log_{10}(A)_c$ using
both estimates shows a decreasing mass across the energy range of the knee region.
For comparison, two models of possible cosmic ray mass variation
near the knee are plotted in Figure \ref{massplot}.
The dotted curve is the model given by Swordy 1995, where a source
spectrum $\propto E^{-3}$ is introduced above the knee region and the energy dependent escape from the Galaxy is assumed
to reach a plateau near the knee region.
 This model provides an
increase in mass across the knee of a size which is larger than for models based simply on particle
rigidity.
The lower dashed line, with which the data seem more consistent, is a model
which has similar low energy behavior as the previous model but which introduces a proton source
which compensates for the lost flux above the SNR acceleration cutoff rigidity,
assumed here to be $10^{15}$V. This is similar to the
suggestion of Protheroe and Szabo 1992\cite{prother92} where an extragalactic component of particles
produced by AGNs provide the cosmic rays at high energy.
These data seem more consistent with this latter model,
where cosmic rays become progressively lighter across the knee region. However, they do not
exclude the possibility that the composition is more or less constant across this energy
range. A sudden change in composition to becoming predominantly iron nuclei ($log_{10}(A)=1.75$)
seems strongly excluded.
The variations in the energy spectrum of Figure \ref{espec} near 8 PeV and the irregularity
in the behavior of '$log_{10}(A)$ rms' near 4PeV shown in Figure \ref{masswidth} lend some support
to the single source model of the knee region discussed by Erlykin and Wolfendale\cite{erl98}.
However, Figure \ref{massplot} does not show the expected increase in $<log_{10}(A)>$ of $\sim$0.4 
across the knee region predicted by this model.

Contrary to conventional wisdom these results do not support a simple `rigidity steepening'
which would lead to a steady increase in mass across the knee region. If the cosmic ray
abundances below the knee are provided with a simple steepening from a spectrum of $E^{-2.75}$
to $E^{-3.0}$ and a composition below the knee is similar to that described by Swordy\cite{sword93},
the size of the effect should be an increase
of $\sim$0.2 in the mean value of $log_{10}(A)$ across the knee. The combination of experimental
measurements discussed here have the sensitivity to detect an increase of this order
if it were present. 

The mass resolution of these methods is poor when compared with direct techniques. The
1$\sigma$ (RMS) widths of the resolution in $log_{10}(A)$ for nuclei have typical size  $\sigma\sim 0.8$.
Hence protons which fluctuate upward by $\sim 2 \sigma$ can be misidentified as iron nuclei.
In this measurement environment any methods which produce a lower intrinsic systematic error in
the technique are extremely valuable. The philosophy behind DICE is to try to reduce these
types of error by decreasing the reliance on Monte Carlo simulations of airshowers in the
interpretation of the data. The method of directly measuring $X_{max}$ from geometry avoids steps
associated with inferences about the location of $X_{max}$ from other variables such as
the density distribution of particles at ground level which are interpreted with
simulations. Only the probability of Cherenkov emission as a function of angle
with the shower axis is needed to estimate $X_{max}$ with DICE. This is relatively insensitive
to the details of the simulation.

\begin{figure}
\vspace{-0.5cm}
\hbox to \textwidth{\hss
\epsfxsize=15cm
\epsfbox{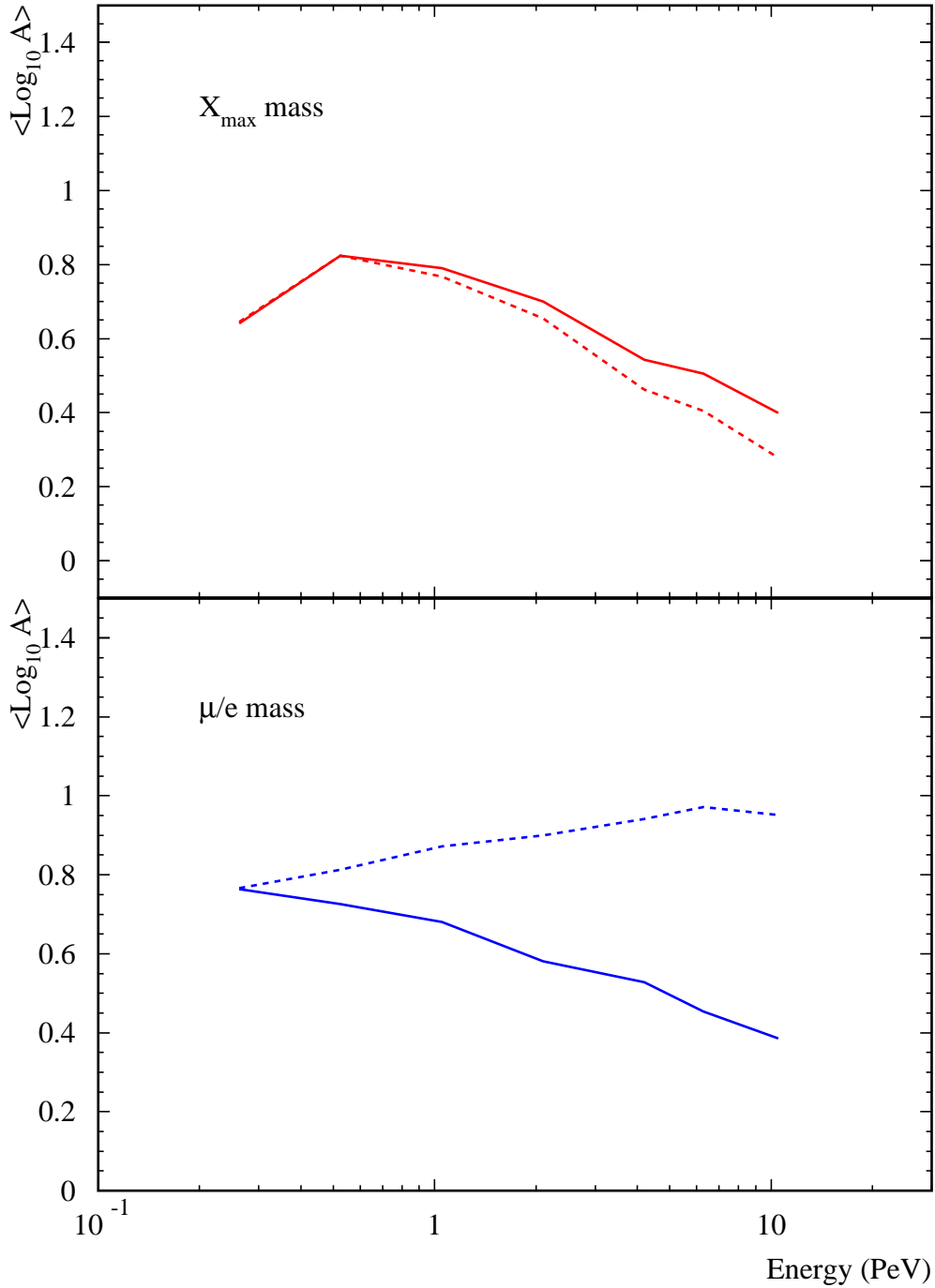}
\hss}
\caption{(Top) The sensitivity of the $X_{max}$ mass measurement to the elongation rate assumed. The solid line
shows the trend of the data with the rate 80$g/cm^2$ as used in this analysis, the dashed line shows the trend
of an analysis using 75$g/cm^2$. (Bottom) The sensitivity of the $\mu/e$ mass measurement to the analysis parameters.
The solid line shows the trend of the data analyzed with a superposition model where $N_\mu\propto E^{0.87}$ as used here.
The dashed line shows the trend produced if the analysis assumes $N_\mu\propto E^{0.82}$.   
\label{sysplot} }
\end{figure}

The use of $X_{max}$ as a variable for composition determination is desirable for other
reasons. The mass estimate derived from this variable is less sensitive to uncertainties in
the characterization of shower parameters.
To explore the sensitivity of the results to the analysis procedures
a simple test can be made on the data set by exploring the trend of the derived masses for
different values of the analysis parameters. The results of this test are shown in Figure \ref{sysplot}.
Here the upper panel shows two lines which are the trend of the data, as shown in Figure \ref{massplot}, under
two assumptions for the value of the elongation rate, $X_{m}$ in equation \ref{eq:xmax}.
The solid line uses $X_{m} = 80 g/cm^2$ in the analysis, the dashed line uses $X_{m} = 75 g/cm^2$.
The smaller elongation rate increases the apparent lightening of composition by an amount of $\sim$0.1
in $<Log_{10}(A)>$, giving an estimate of the systematic uncertainties in the analysis of the DICE $X_{max}$ data.
The lower panel shows a similar test of the $\mu/e$ mass derived from ground particle densities.
Here the solid line represents the superposition model where $\alpha = 0.87$ in equation \ref{eq:muemass},
the dashed line is an analysis where $\alpha = 0.82$. This mass estimate is clearly sensitive
to the precise value of $\alpha$ used since this change makes $<Log_{10}(A)>$ apparently {\it increase}
by an amount $\sim$0.2 across this energy region. The systematic error in the $\mu/e$ mass is therefore large
since simulations typically give values of $\alpha$ in the range $\sim 0.8-0.9$. 
The above analyses illustrate a crucial property of multiparameter measurements in this type of work: the
requirement for consistency from a variety of apparently redundant measurements can directly assess the validity
of the result. In the case discussed here the mean mass estimates from the different techniques must agree
since these are the same events, consequently an analysis where $X_{m} = 75g/cm^2$ and $\alpha = 0.82$
is excluded by this consistency check. 

The instrumentation of ground based cosmic-ray air shower detectors have improved significantly over the
past decade or so. The statistical quality of these measurements has improved to the point where the intrinsic fluctuations
in the shower developments are becoming the limiting factor in the determination of primary composition.
The new challenge emerging in the derivation of composition from measurements of airshowers is the
identification and quantification of systematic errors in aspects of the measurements themselves or in
the simulations used for analysis. By making use of multiparameter measurements of the same showers
we can directly explore our level of understanding of these issues and provide a reliable determination
of cosmic-ray composition in this energy region. 
The results presented here are an attempt to take some initial steps
in this direction.

\begin{ack}
The authors gratefully acknowledge the assistance of the HiRes collaboration and Ryan Lawrence.
We thank M. Cassiday for technical support and the commander and staff of the US Army Dugway Proving
Grounds for assistance.
Thanks are due to our colleagues who collected much of these data, these include Kevin Boothby,
Mark Chantell, Curtis Larsen, Kevin Green, and the entire CASA-MIA group. We are also grateful to useful discussions
with Clem Pryke.
This work is supported through the Block Fund at the University of Chicago and institutional support
and by the National Science Foundation Grant PHY 9514193 at the University of Utah.
\end{ack}

\newpage
\appendix
\begin{center}
{\bf Appendices}
\section{Appendix of Fitting Functions}
\end{center}
\subsection{Cherenkov Light Distribution Function}
This function which is used to fit the Cherenkov light sizes at the DICE detectors is derived from
simulations using CORSIKA 4.5 with VENUS \cite{corsika98}
 and is valid for shower core distances larger than
75m from each DICE cluster. It depends on particle energy, mass and event geometry as follows:
\begin{equation}
Ch=C_0 E(PeV)^\gamma A^{-\epsilon} e^{-\beta r(m)} \label{eq:csize}
\end{equation}
where
\begin{eqnarray*}
C_0 & = & 1.89\times 10^6 \\
\gamma & = & 1.144+0.0905\times log_{10}(A) \\
\beta  & = & 0.0161\times(1-0.128\times  log_{10}(A) \\
 & & + 0.124 \times  log_{10}(E(PeV))\times (1+0.322\times  log_{10}(A)))  \\
\epsilon & = & 0.186  \\
\end{eqnarray*}

Here $Ch$ is the number of Cherenkov photons collected in the effective area (3.75m$^2$) of a DICE detector.
The total energy of the shower in PeV is $E(PeV)$, A is the mass of the incident particle, 
and $r(m)$ is the perpendicular distance to the shower core
from the DICE detector. Note the amount of light varies almost linearly with energy ($\gamma$) since the
Cherenkov size is close to a calorimeteric measurement of the shower energy because nearly all of the particles
produced in the shower give Cherenkov light.
The light intensity also decreases with increasing mass for the same total energy ($\epsilon$).
The light intensity falls with distance from the shower
core, for
low mass particles near 1PeV the light intensity falls with a scale distance of $\sim$62m ($1/\beta$).
For iron nuclei at similar energies the scale distance is $\sim$80m.

\subsection{$X_{max}$ Fitting Function}
The following equation is the fitting function used for the primary mass and energy dependence for
the location of $X_{max}$:

\begin{equation}
X_{max} = X_0 + X_m \times ( log_{10}(E) - log_{10}(A)) \label{eq:xmax}
\end{equation}
where
\begin{eqnarray*}
X_0 & = & 560 g/cm^2 \\
X_m & = & 80 g/cm^2 \\
\end{eqnarray*}

Where $X_{max}$ is the location of shower maximum, E and A are  
the incident particle energy in PeV and particle mass (in nuclear mass units) respectively.

\subsection{Gaisser Shower Size Function}
The function used for the average shower development is adapted from Gaisser \cite{gaisser}
to calculate the expected electron size at ground level, $X_0$, for a shower of total energy $E$ with
a maximum at a depth $X_{max}$:

\begin{equation}
N_e(X_0)=A_e\times E(PeV)\times p/(p+1)e^p(X_0/(X_{max}-\lambda))^{(p+1)}e^{-X_0/\lambda} \label{eq:gaiss}
\end{equation}

Here $A_e = 6.02\times10^5$ is a constant, $E(PeV)$ is the shower energy in PeV, $X_0=860g/cm^2$ is the
average detector atmospheric depth, $p=X_{max}/\lambda-1$ and $\lambda=70g/cm^2$ is the hadronic interaction length.
This assumes simple scaling in the Gaisser function, i.e. $S_0=0.045$ and $\epsilon=0.074$ GeV.
The lower right panel of Figure \ref{gfig} shows the effect of the scaling violation suggested by Gaisser\cite{gaisser}
as a dashed curve.

\clearpage

\begin {center}
\section{Appendix of Tabulated Results}
This Appendix contains tables of the results of this investigation.
\end{center}

\begin{table}[h]
\label{tab:xmax}
\caption{Table of $X_{max}$ versus energy from DICE}
\begin{tabular}{c c c}\hline
Energy Bin & Median Energy & $<X_{max}>$ \\
(PeV)  & (PeV)  & $g/cm^2$  \\ \hline
0.2-0.4 &      0.26  $\pm$      0.05 &      470.7   $\pm$      0.6 \\
0.4-0.7 &      0.50  $\pm$      0.10 &      476.3  $\pm$       0.8 \\
0.7-1.5 &      0.91  $\pm$      0.18 &      497.7  $\pm$       1.0 \\
1.5-3.0 &      1.92 $\pm$       0.38 &      529.6  $\pm$       1.6 \\
3.0-5.0 &     3.66  $\pm$      0.73 &      562.8   $\pm$      3.3 \\
5.0-8.0 &      6.03  $\pm$      1.21 &      587.4   $\pm$      5.5 \\
8.0-15.0 &     10.09 $\pm$       2.02 &      610.3 $\pm$       10.0 \\
\hline
\end{tabular}
\end{table}

\begin{table}
\label{tab:mass}
\caption{Table of mass estimates from DICE, ground sizes from CASA/MIA and from a combination of the two methods.
Also given is the result of the (p+$\alpha$)/all fraction estimate, discussed in the text.} 
\begin{tabular}{c c c c c}\hline
Energy & Combined & $X_{max}$ & $\mu/e$ & (p+$\alpha$)/all \\
(PeV) & $<log_{10}(A)>$ &  $<log_{10}(A)>$ & $<log_{10}(A)>$ &  \\ \hline
       0.2 &       0.61  $\pm$    0.01 &       0.64  $\pm$    0.01 &       0.76   $\pm$   0.01 & 0.609 $\pm$    0.011  \\
       0.5 &       0.71  $\pm$    0.01 &       0.82  $\pm$    0.01 &       0.73  $\pm$    0.02 & 0.560 $\pm$    0.012 \\
       1.0 &       0.70  $\pm$    0.01 &       0.79  $\pm$    0.01 &       0.68  $\pm$    0.02 & 0.531 $\pm$    0.014 \\
       2.0 &       0.63  $\pm$    0.02 &       0.70  $\pm$    0.02 &       0.58  $\pm$    0.03 & 0.494 $\pm$    0.022  \\
       4.0 &       0.53  $\pm$    0.04 &       0.54  $\pm$    0.04 &       0.53  $\pm$    0.05 & 0.577 $\pm$    0.045  \\
       6.0 &       0.47  $\pm$    0.06 &       0.51  $\pm$    0.07 &       0.45  $\pm$    0.09 & 0.633 $\pm$    0.082 \\
      10.0 &       0.34  $\pm$    0.10 &       0.40  $\pm$    0.12 &       0.39  $\pm$    0.12 & 0.674 $\pm$    0.112 \\
\hline
\end{tabular}
\end{table}

\begin{table}
\label{tab:fluxes}
\caption{The All Particle Fluxes of Cosmic Rays from DICE}
\begin{tabular}{ c c c}\hline
Total Particle Energy (GeV) & Flux (m$^2$ sr s GeV)$^{-1}$ & Flux Error ($\pm$(m$^2$ sr s GeV)$^{-1}$) \\ \hline
 0.355$\times 10^6$ &   0.361$\times 10^{-10}$ &   0.50$\times 10^{-12}$ \\ 
 0.447$\times 10^6$ &   0.208$\times 10^{-10}$ &   0.30$\times 10^{-12}$ \\ 
 0.562$\times 10^6$ &   0.115$\times 10^{-10}$ &   0.18$\times 10^{-12}$ \\ 
 0.708$\times 10^6$ &   0.619$\times 10^{-11}$ &   0.11$\times 10^{-12}$ \\ 
 0.891$\times 10^6$ &   0.322$\times 10^{-11}$ &   0.65$\times 10^{-13}$ \\ 
 0.112$\times 10^7$ &   0.167$\times 10^{-11}$ &   0.39$\times 10^{-13}$ \\ 
 0.141$\times 10^7$ &   0.906$\times 10^{-12}$ &   0.24$\times 10^{-13}$ \\ 
 0.178$\times 10^7$ &   0.496$\times 10^{-12}$ &   0.15$\times 10^{-13}$ \\ 
 0.224$\times 10^7$ &   0.263$\times 10^{-12}$ &   0.96$\times 10^{-14}$ \\ 
 0.282$\times 10^7$ &   0.142$\times 10^{-12}$ &   0.61$\times 10^{-14}$ \\ 
 0.355$\times 10^7$ &   0.734$\times 10^{-13}$ &   0.38$\times 10^{-14}$ \\ 
 0.447$\times 10^7$ &   0.333$\times 10^{-13}$ &   0.23$\times 10^{-14}$ \\ 
 0.562$\times 10^7$ &   0.163$\times 10^{-13}$ &   0.14$\times 10^{-14}$ \\ 
 0.708$\times 10^7$ &   0.711$\times 10^{-14}$ &   0.82$\times 10^{-15}$ \\ 
 0.891$\times 10^7$ &   0.557$\times 10^{-14}$ &   0.64$\times 10^{-15}$ \\ 
 0.112$\times 10^8$ &   0.207$\times 10^{-14}$ &   0.35$\times 10^{-15}$ \\ 
 0.141$\times 10^8$ &  0.999$\times 10^{-15}$ &   0.21$\times 10^{-15}$ \\ 
 0.178$\times 10^8$ &   0.430$\times 10^{-15}$ &   0.12$\times 10^{-15}$ \\ 
 0.224$\times 10^8$ &   0.113$\times 10^{-15}$ &   0.57$\times 10^{-16}$ \\ 
 0.282$\times 10^8$ &   0.112$\times 10^{-15}$ &   0.50$\times 10^{-16}$ \\ 
\hline
\end{tabular}
\end{table}

\clearpage

\begin{center}
\section{Appendix detailing the muon mass function}
\end{center}

\begin{figure}[b]
\vspace{-0.5cm}
\hbox to \textwidth{\hss
\epsfxsize=15cm
\epsfbox{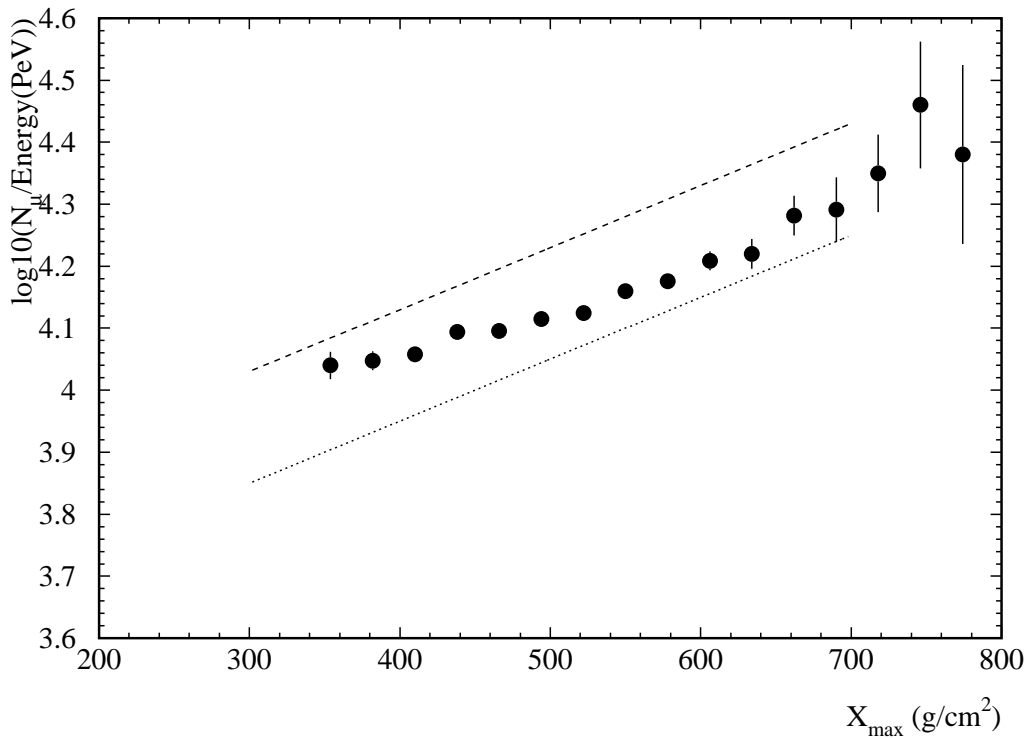}
\hss}
\caption{The variation of the normalized log10(muon size/energy) with depth of shower maximum ($X_{max}$)
around 1 PeV is shown as the filled symbols. There is a clear correlation between the depth of shower maximum
and the normalized muon size. The lower dotted line is a simulation of this effect for masses $A\le 4$, the
upper dashed line is a simulation for $A > 4$ using the same composition discussed in the text.
\label{muxmx} }
\end{figure}

As has been mentioned in the main text the mass derived from the muon size, equation \ref{eq:muemass},
is susceptible to variations introduced by fluctuations from the location of shower maximum. Although
these are not as large as associated with the electron size shown in Figure \ref{gfig}, they produce
poor resolution in the muon mass function unless corrected. A quantitative measure of this effect is shown in 
Figure \ref{muxmx}

The normalized muon mass size ($log_{10}(N_\mu/E(PeV)$) is derived from the MIA muon size and the DICE Cherenkov
energy, E(PeV).
This size is clearly correlated with $X_{max}$ depth at energies near
1 PeV as shown. This is a weaker version of the effect shown for electrons in Figure \ref{gfig}. It presumably
arises from more muons either decaying or reaching the end of the range for the showers with maxima higher from
the ground. 
Although at a fixed $X_{max}$ the muon mass function will perform as desired, the presence of fluctuations smears
the possible mass resolution because of this effect. Indeed as the relationship between the simulation curves
(dashed lines) on Figure \ref{muxmx} and the measured data show, in general as the shower penetrates more
deeply the muon content appears more `proton-like' as expected. To produce a more accurate mass function
a variable must be introduced to offset the effect of shower depth shown in the Figure. Since electron size
is closely correlated with $X_{max}$, as shown in Figure \ref{gfig}, this can be used to make a correction
and provide a mass function from the ground sizes which is completely independent from the $X_{max}$ measurement
of DICE and therefore provide a separate mass estimate for each shower.

Figure \ref{muxmx} shows the basis for this correction. For showers near 1 PeV the average muon size, normalized
by the Cherenkov energy, is plotted against the normalized electron size. The variations in these sizes come predominantly
from the fluctuations in $X_{max}$ between showers.  Figure \ref{muxmx} represents on average how the muon size
changes with electron size due to these fluctuations. The dashed line is a linear fit to these data with a slope
of 0.5. The electron size correction to the muon size mass measure is therefore of the form:

\begin{equation}
\delta (e_g)= 0.5\times (e_g-1.14\times log_{10}E(PeV))  \label{eq:delta}
\end{equation}

Where $e_g=log_{10}(N_{ec}/N_{e0})$ as before, and where the second term corrects for the average electron size
increase with energy. This effectively provides a correction for the muon size growth with $X_{max}$ shown
in Figure \ref{muxmx}.

\begin{figure}
\vspace{-0.5cm}
\hbox to \textwidth{\hss
\epsfxsize=15cm
\epsfbox{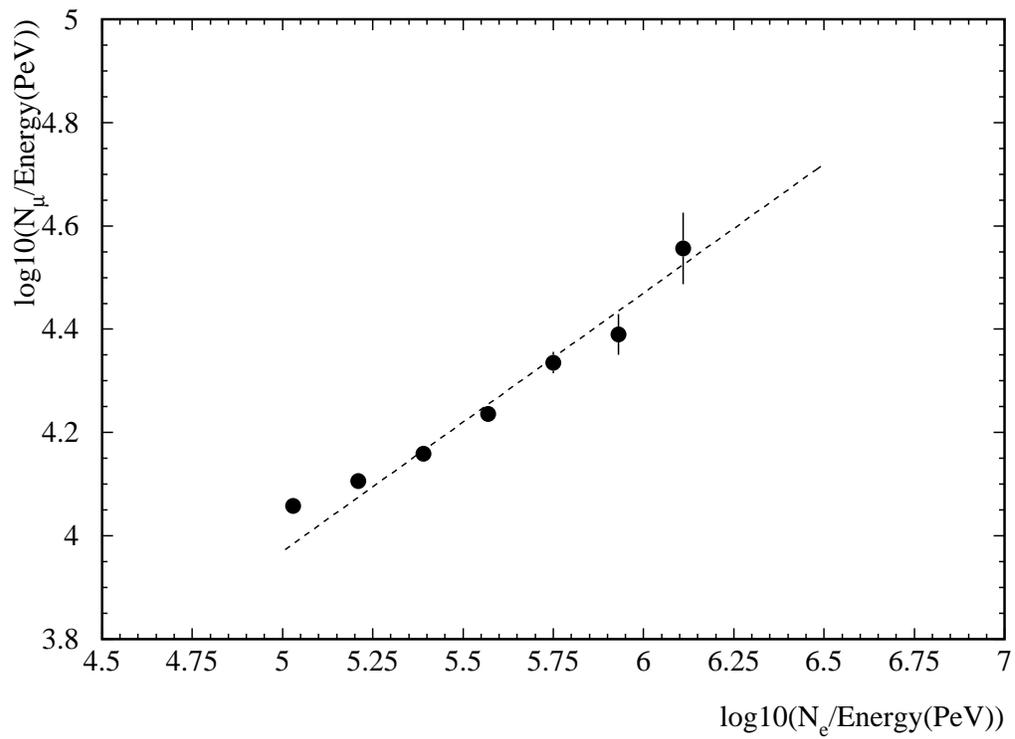}
\hss}
\caption{The variation of the normalized log10(muon size/energy) with normalized log10(electron size/energy)
near 1 PeV. The line is a fit with a slope of 0.5, this is discussed further in the text
\label{muel} }
\end{figure}

\end{document}